
\magnification\magstep1
\baselineskip = 0.5 true cm
\parskip=0.1 true cm

  \def\sa{\vskip 0.30 true cm}
  \def\sb{\vskip 0.60 true cm}
  \def\sc{\vskip 0.15 true cm}

  \nopagenumbers

  \vsize = 22 true cm
  \hsize = 15 true cm


\font\msim=msym10
\def\gr{\hbox{\msim R}}

\def\grn{\hbox{\msim N}}
\def\grz{\hbox{\msim Z}}
\def\grc{\hbox{\msim C}}


   \line{\vbox{\hsize 4.4 true cm
   \noindent
P}\hfill \vbox{\hsize 2.8 true cm
   \noindent
   \bf LYCEN 9234\break
   Sept. 1992}}

\sc
\sb
\sc
\sb

\centerline {\bf INTRODUCTION TO QUANTUM ALGEBRAS$^*$}

\sa
\sb
\vskip 0.5 true cm

\centerline {Maurice R.~Kibler}

\sa

\centerline {Institut de Physique Nucl\'eaire de Lyon}
\centerline {IN2P3-CNRS et Universit\'e Claude Bernard}
\centerline {43 Boulevard du 11 Novembre 1918}
\centerline {F-69622 Villeurbanne Cedex, France}

\sa
\sa
\sa
\sb
\sb

\baselineskip = 0.7 true cm

\sa

\centerline {\bf Abstract}

\sb

The concept of a quantum algebra is made easy through the
investigation of the prototype algebras $u_{qp}(2)$, $su_q(2)$
and $u_{qp}(1,1)$. The latter quantum algebras are introduced
as deformations of the corresponding Lie algebras~; this
is achieved in a simple
way by means of $qp$-bosons. The Hopf algebraic structure of
$u_{qp}(2)$ is also discussed. The basic
ingredients for the representation theory of $u_{qp}(2)$ are
given. Finally, in connection with the quantum
algebra $u_{qp}(2)$, we discuss the $qp$-analogues of the harmonic
oscillator and of the (spherical and hyperbolical) angular
momenta.

\sa
\sa
\sa
\sb

\baselineskip = 0.5 true cm
\noindent $^*$ Lectures presented at The Second International
School of Theoretical Physics ``Symmetry and Structural
Properties of Condensed Matter'', Pozna\'n, Poland (26 August~-~2
September 1992). Published in ``Symmetry and Structural
Properties of Condensed Matter'', W. Florek,
                                  D. Lipi\'nski,
                              and T. Lulek, Eds.~(World
Scientific, Singapore, 1993). pp.~445-464.

\vfill\eject

\vglue 0.8 true cm

\centerline {\bf INTRODUCTION TO QUANTUM ALGEBRAS}

\sa
\sb
\vskip 0.5 true cm

\centerline {Maurice R.~Kibler}

\sa
\baselineskip = 0.50 true cm

\centerline {Institut de Physique Nucl\'eaire de Lyon}
\centerline {IN2P3-CNRS et Universit\'e Claude Bernard}
\centerline {43 Boulevard du 11 Novembre 1918}
\centerline {F-69622 Villeurbanne Cedex, France}

\sa
\sa
\sa
\sb
\sb
\sa

\centerline {\bf Abstract}

\sb

\baselineskip = 0.48 true cm
\leftskip = 1.0 true cm
\rightskip = 1.0 true cm

The concept of a quantum algebra is made easy through the
investigation of the prototype algebras $u_{qp}(2)$, $su_q(2)$
and $u_{qp}(1,1)$. The latter quantum algebras are introduced
as deformations of the corresponding Lie algebras~; this
is achieved in a simple
way by means of $qp$-bosons. The Hopf algebraic structure of
$u_{qp}(2)$ is also discussed. The basic
ingredients for the representation theory of $u_{qp}(2)$ are
given. Finally, in connection with the quantum
algebra $u_{qp}(2)$, we discuss the $qp$-analogues of the harmonic
oscillator and of the (spherical and hyperbolical) angular
momenta.

\sa
\sb

\leftskip = 0 true cm
\rightskip = 0 true cm
\baselineskip = 0.59 true cm

\centerline {\bf 1. Where Does It Come From~?}

The notion of deformation is very familiar to the physicist.
In this connection, quantum mechanics may be considered as a
deformation (the deformation parameter being $\hbar$) of
classical mechanics and relativistic mechanics is, to a certain
extent, another deformation (with $1/c$ as deformation
parameter) of classical mechanics. Although a sharp distinction
should be established between {\it deformations} and {\it
quantized universal enveloping algebras} or {\it quantum
algebras}, the concept of a quantum algebra is more easily
introduced in the parlance of deformations. Along this vein,
the idea of deformed bosons, introduced as early as in the seventies [1,2,5],
plays an important role.

The concept of a quantum algebra (or quantum group) goes back
to the end of the seventies. It was introduced, under different
names, by Kulish, Reshetikhin, Sklyanin, Drinfeld (from the
Faddeev school) and Jimbo [3,4,6,7] in terms of a {\it quantized
universal enveloping algebra} or an {\it Hopf bi-algebra} and,
independently, by Woronowicz [8] in terms of a {\it compact
matrix pseudo-group}.

Among the various motivations that led to the concept of a
quantum group, we have to mention the quantum
inverse scattering technique, the solution of the quantum Yang-Baxter
equation and, more generally, the study of exactly solvable models in
statistical mechanics. More recent applications of quantum
algebras concern~: conformal field theories in two dimensions~;
(quantum) dynamical
systems~; quantum optics~; molecular, atomic and nuclear
spectroscopies~; condensed matter physics~; knot theory, theory
of link invariants (Jones polynomials) and braid
groups~; and so on. In addition, the
concept of a quantum group constitutes a basic tool in
non-commutative geometry. Thus, quantum groups are of paramount
importance not only in physics and quantum chemistry but in
pure mathematics equally well.

It is the aim of this series of lectures to give a primer on
quantum algebras. The lectures are organized as follows. Some
$qp$-deformed bosons and a $qp$-deformed harmonic oscillator are
introduced in \S 2 and 3. In \S 4, representations other than the
Fock representation are given for the deformed boson algebra.
Section 5 deals with the quantum algebras $u_{qp}(2)$,
$su_{q}(2)$ and $u_{qp}(1,1)$ while the representation theory
and the Hopf algebraic structure of $u_{qp}(2)$ are considered
in \S 6. Section 7 is devoted to an (incomplete)
classification of the possible applications of quantum
algebras. Finally, a sketch of bibliography is given is \S 8.

\sb

\centerline {\bf 2. Introducing $qp$-Bosons}

We start with the usual (one-particle) Fock space
$$
{\cal F} = \left \{ |n> \; \; : \; n \in \grn \right \}
\eqno (1)
$$
which is very familiar to the physicist. The state vectors
$\vert n >$ are the eigenstates for an ordinary harmonic
oscillator in one dimension. In the terminology to
be used below, ${\cal F}$ is a {\it non-deformed} Fock space.

Definition 1. Let us define the linear operators $a^+$, $a$ and
$N$ on the vector space ${\cal F}$ by the relations
$$
a^+ \; |n> \; = {\sqrt {[n + 1]}} \; |n + 1 > \quad
a   \; |n> \; = {\sqrt {[n]    }} \; |n - 1 > \quad
N   \; |n> \; = n \; |n     >
\eqno (2)
$$
with $a \; | 0> \; = 0$, where we use the notation
$$
[c] \; \equiv \; [c]_{qp} \; = \; { {q^c - p^c} \over {q - p} }
\qquad (c \in \grc),
\eqno (3)
$$
the two parameters $p$ and $q$ being fixed parameters
taken (a priori) in the field of complex numbers \grc.

It is to be observed that in the limiting
case $p=q^{-1} \to 1$, we have simply $[c]=c$
so that $a^+$, $a$ and $N$ are (respectively) in this case the
ordinary creation, annihilation and number operators
occurring in various areas of theoretical
physics. In the other cases, the operators
$a^+$ and $a$ defined by equations (1-3) are called $qp$-deformed
creation and annihilation operators, respectively, and
they are collectively referred to as $qp$-deformed
bosons or simply $qp$-bosons. Observe that the operator
$N$ is a non-deformed operator that coincides with the usual
number operator.

The (complex) number $[c]$ defined by (3) is a
$qp$-deformed number. It can be rewritten as
$$
[c] = { \sinh (c {s-r \over 2}) \over \sinh ({s-r \over 2}) }
\> \exp \left[ (c-1) {s+r \over 2} \right]
\eqno (4)
$$
where
$$
s = \ln q \qquad r = \ln p
\eqno (5)
$$
(Some algebraic relations satisfied by such
$qp$-deformed numbers are listed in the Appendix.)
Two particular situations are of special interest, viz., $p =
q^{-1}$ ($r=-s$) and $p=1$ ($r=0$). The situation $p=q^{-1}$,
for which
$$
[c] = {{q^c - q^{-c}} \over {q - q^{-1}}}
    = {{\sinh(c \ln q)} \over {\sinh(\ln q)}},
\eqno (6)
$$
is
mainly encountered in the physical literature while the
situation $p=1$, for which
$$
[c] = {{q^c - 1} \over {q - 1}}
    = q^{{c-1} \over 2} \>
{\sinh ({c \over 2} \ln q) \over
 \sinh ({1 \over 2} \ln q)},
\eqno (7)
$$
comes from the mathematical literature. Observe that for
$c \in \gr$, the numbers $[c]$ given by (6) are real for
$q \in \gr$ or $q \in S^1$.

Note that a simple iteration of equation (2) yields
$$
\vert n> = { 1 \over \sqrt{[n]!} } \> (a^+)^n \> \vert 0>
\eqno (8)
$$
where the $qp$-deformed factorial $[n]!$ is defined by
$$
[n]! = [n] [n-1] \cdots [1]
\eqno (9)
$$
for $n \in \grn$ with $[0]! = 1$.

Property 1. As a trivial property, we have
$$
(a)^{\dagger} = a^+ \qquad (N)^{\dagger} = N \qquad
[N,a^+] = a^+ \qquad [N,a] = -a
\eqno (10)
$$
where $(X)^{\dagger}$ denotes the adjoint of the operator $X$ and
$[X,Y] \equiv [X,Y]_- = XY -YX$ the commutator of $X$ and $Y$.
Equation $(a)^{\dagger} = a^+$ is valid under the condition
(which is supposed to hold in this paper) that the
$qp$-deformed numbers $[n]$, $n \in \grn$, are real~; this is
certainly the case if
$q          \in \gr$ and
$p          \in \gr$ or if
$q = p^{-1} \in S^1$.\footnote* {\sevenrm
Note that for $q = p^{-1} = \exp (i \varphi)$,
with $\varphi \in \gr$, the $q$-deformed number $[n+1]$, with
$n \in \grn$, is nothing but the Chebyshev polynomial of the
second kind $U_n(\cos \varphi)$.}

Property 2. As a basic property, we can check that
$$
a a^+ = [N + 1] \qquad
a^+ a = [N]
\eqno (11)
$$
where we use the abbreviation
$$
[X] \; \equiv \; [X]_{qp} \; = \; { {q^X - p^X} \over {q - p} }
\qquad X \in {\cal F}
\eqno (12)
$$
which parallels for operators the defining relation (3) for
numbers.

As a corollary of Property 2, we can derive the following
expression for the $Q$-mutator $[a,a^+]_Q$~:
$$
[a,a^+]_Q \equiv aa^+ - Qa^+a = {1 \over {q-p}}
[q^N (q-Q) - p^N (p-Q)]
\eqno (13)
$$
where $Q$ may be any complex number.
(The $Q$-mutator reduces to the
ordinary commutator when $Q=1$.)
Three special cases are of
importance. First, for $p=1$ we have
$$
[a, a^+]_q = 1 \qquad [a, a^+]_1 = q^N
\eqno (14)
$$
that correspond to the deformed bosons
encountered mainly in the mathematical literature. (In passing,
note that the relation $[a, a^+]_q = 1$, which interpolates between
fermions and bosons for $-1 \le q \le 1$, may be of interest in
anyonic statistics.) Second, for $p=q^{-1}$ we have
$$
[a, a^+]_q        = q^{-N} \quad
[a, a^+]_{q^{-1}} = q^N    \quad
[a, a^+]_1        = {1 \over {q - q^{-1}}} [q^N(q-1) - q^{-N}(q^{-1}-1)]
\eqno (15)
$$
that correspond to the deformed bosons introduced in the
physical literature. Third, for generic $p$ and $q$ we have
$$
[a, a^+]_q = p^N \qquad
[a, a^+]_p = q^N \qquad
[a, a^+]_1 = {1 \over {q - p}} [q^N(q-1) - p^N(p-1)]
\eqno (16)
$$
that subsume (14) and (15) and
that lead to the usual commutation relations
for ordinary bosons (corresponding to $p = q^{-1} \to 1$).

At this point, it is worth to note that any set $\left \{ a, a^+\right \}$
of $qp$-bosons acting on the Fock space ${\cal F}$
can be connected to any set $\left \{ b, b^+\right \}$ of
$hf$-bosons acting on the same space ${\cal F}$. Indeed, we have
$$
a   = s(N+1) \> b   = b \> s(N) \qquad
a^+ = b^+ \> s(N+1) = s(N) \> b^+
\eqno (17)
$$
where the (operator-valued) scaling factor $s$ is given by
$$
s(N) = \left( { [N]_{qp} \over [N]_{hf} } \right)^{1 \over 2}
\eqno (18)
$$
Of course, the passage expressions (17-18) require that none
of the involved $[N]$-operators be vanishing. This excludes, in
particular, that $q$ be a root of unity when $p=q^{-1}$. As a
particular case, any set $\left \{ a, a^+\right \}$ of
$qp$-bosons can be related through (17-18) to a set
$\left \{ b, b^+\right \}$ of ordinary bosons corresponding to
the limiting case $f = h^{-1} \to 1$. As another particular
case, the set of $qp$-bosons $\left \{ a, a^+\right \}$
with $p = q^{-1}$ can be connected to the set of $hf$-bosons
$\left \{ b, b^+\right \}$ with $f=1$ and $h = q^2$ through the
relations
$$
b = q ^ { {N \over 2} } \> a = a \> q ^ { {N-1} \over 2 }
\qquad
b^+ = a^+ \> q^{N \over 2} = q^{{N-1} \over 2} \> a^+
\eqno (19)
$$
which follow from (17-18)~; equation $(19)$ thus
allows us to pass from the deformed bosons satisfying
$$
a a^+ - q      a^+ a = q^{-N} \qquad
a a^+ - q^{-1} a^+ a = q^N
\eqno (20)
$$
to the deformed bosons satisfying
$$
b b^+ - q^2 b^+ b = 1
\eqno (21)
$$
Finally,
note that equations (17-18) make it possible to connect the set (of
ordinary bosons) $\left\{ b, b^+ \right\}$ with $f = h^{-1} \to 1$
satisfying $ b b^+ - b^+ b = 1 $ to the set (of fermions)
$\left\{ a, a^+ \right\}$ with $p = - q \to 1$ satisfying
$ a a^+ + a^+ a = 1 $.

Going back to the more general $qp$-bosons,
we give a few elements useful for
constructing  $qp$-deformed coherent states. First, let us make
the replacements
$$
\vert n> \mapsto {1 \over \sqrt {[n]!}} z^n \qquad
a^+      \mapsto z                          \qquad
a        \mapsto D_{qp}                     \qquad
N        \mapsto z { \partial \over {\partial z} }
\eqno (22)
$$
where $z \in \grc$ and $D_{qp}$ is the finite difference
operator defined via its action on any function $f(x)$ as
$$
D_{qp} f(x) = { {f(qx) - f(px)} \over {(q-p)x} }
\eqno (23)
$$
(For evident reasons, the operator $D_{qp}$ may be called
$qp$-derivative. Some information on $D_{qp}$ may be found
in the Appendix.) It is easy to check that equation (22)
defines a realization (a $qp$-deformed Bargmann
realization, indeed) of the basic relations (2) defining the
number operator $N$ and the $qp$-bosons $a$ and $a^+$. Second,
let us introduce in ${\cal F}$ the state vector
$$
\vert z> = \sum_{n=0}^{+ \infty}
{ 1 \over \sqrt {[n]!} } \> z^n \> \vert n>
\eqno (24)
$$
It can be checked that the eigenvalue equation
$$
a \> \vert z> = z \> \vert z>
\eqno (25)
$$
holds in the space ${\cal F}$. Therefore, the state $\vert z>$
can be considered as a (non-normalized) $qp$-deformed coherent
state. The normalized $qp$-deformed coherent state reads
$$
\vert z> = \left( \exp_{qp} (\vert z \vert^2) \right)^{-{1 \over 2}}
\sum_{n=0}^{+ \infty} { 1 \over \sqrt {[n]!} } \> z^n \> \vert n>
\eqno (26)
$$
where the $qp$-deformed exponential $\exp_{qp}$ is defined by
$$
\exp_{qp} (x) = \sum_{n=0}^{+ \infty} {x^n \over [n]!} \>
\eqno (27)
$$
(See the Appendix for a property of the $qp$-exponential.)
We stop here our discussion around $qp$-deformed coherent
states. (A complete study requires the introduction of a
measure on the $qp$-deformed Bargmann space
$\left \{ z^n / \sqrt {[n]!} \; \; : \; n \in \grn \right \}$.)

\sb

\centerline {\bf 3. A $qp$-Analogue of the Harmonic Oscillator}

We are now in a position to introduce a $qp$-deformed harmonic
oscillator. The literature on this subject is now abundant and
the reader may consult, for example, Refs.~[5,10,11] for further
details,
especially for the particular situations where $p=q^{-1}$ and
$p=1$.

Definition 2. From the $qp$-deformed creation and annihilation
operators $a$ and $a^+$, let us define the operators
$$
p_x \; = \; i  \; \sqrt{ {\hbar \mu \omega} \over 2 }  \; (a^+ - a)
\qquad
  x \; =       \; \sqrt{ \hbar \over {2 \mu \omega} }  \; (a^+ + a)
\eqno (28)
$$
acting on ${\cal F}$,
where $\hbar$, $\mu$ and $\omega$ have their usual meaning in
the context of the (ordinary) harmonic oscillator.

Equation (28) defines $qp$-deformed momentum and position operators
$p_x$ and $x$, respectively, and bears the same form as for the ordinary
creation and annihilation operators corresponding to the
limiting case $p = q^{-1} \to 1$.

Property 3. The commutator of the $qp$-deformed
operators $x$ and $p_x$ is
$$
[x, p_x] \; = \; i \hbar \; [ a, a^+]
         \; = \; i \hbar \; ([N+1] - [N])
         \; = \; i \hbar \; {1 \over {q-p}} [q^n(q-1) - p^n(p-1)]
\eqno (29)
$$
which reduces to the ordinary value $i\hbar$ in the limiting
case $p = q^{-1} \to 1$.

Thus, we may think of a $qp$-deformed ($N$-dependent)
uncertainty principle. In particular for $p = q^{-1}$, equation
(29) can be specialized as
$$
[x, p_x] \; = \; i \hbar \;
{\cosh [(n + {1 \over 2}) \ln q] \over \cosh ({1 \over 2} \ln q)}
\eqno (30)
$$
in terms of eigenvalues. The right-hand side of (30)
increases with $n$ (i.e., with the energy, see equation (33) below) and is
minimum as well as $n$-independent in the limiting case $q = 1$ [11].

Definition 3. We define the self-adjoint operator $H$ on ${\cal F}$
by
$$
H = {1 \over {2\mu}} \, {p_x}^2 + {1 \over 2} \, \mu \, \omega^2
\, x^2 = {1 \over 2} \, (a^+a \, + \, aa^+) \, \hbar \, \omega
= {1 \over 2} \left( [N]  +  [N+1] \right) \, \hbar \, \omega
\eqno (31)
$$
in terms of the $qp$-deformed operators previously defined.

In the limiting case $p = q^{-1} \to 1$,
the operator $H$ is nothing but the
Hamiltonian for a one-dimensional harmonic oscillator.
We take
equation (31) as the defining relation for a $qp$-deformed
one-dimensional harmonic oscillator. The case of a $qp$-deformed
$d$-dimensional, with $d \ge 2$, (isotropic or anisotropic)
harmonic oscillator can be handled from a superposition of
one-dimensional $qp$-deformed oscillators.

Property 4. The spectrum of $H$ is given by
$$
E \; \equiv \; E_n \; = \; {1 \over 2} \; ([n] + [n+1]) \; \hbar \; \omega
                      = \; {1 \over 2} \;
{{q^n(q+1) - p^n(p+1)} \over {q-p}}                     \; \hbar \; \omega
\qquad n \in \grn
\eqno (32)
$$
and is discrete.

This spectrum turns out to be a deformation of the one for the
ordinary one-dimensional harmonic oscillator corresponding to
the limiting case $p = q^{-1} \to 1$. The levels are shifted (except the
ground level) when we pass from the ordinary harmonic
oscillator to the $qp$-deformed harmonic oscillator~:
the levels are not uniformly spaced.

In the special case $p=q^{-1}$, equation (32) yields
$$
E_n \; = \; {1 \over 2} \;
{\sinh [(n + {1 \over 2}) \ln q] \over \sinh ({1 \over 2} \ln q)}
\; \hbar \omega
\eqno (33)
$$
and, as expected, $E_n$ is real for $q \in S^1$
($q = \exp (i \varphi)$ with $\varphi \in \gr$) or
$q \in \gr$. In this case, we have
$$
E_{n+1} - E_{n} \; = \; \hbar \omega \; \cosh [(n+1) \ln q]
\eqno (34)
$$
which is $n$-independent only for $q=1$.

At this place, it is interesting to mention that we may think
to obtain the $qp$-deformed spectrum (32) from the
$qp$-deformed Schr\"odinger equation (by using the
$qp$-derivative $D_{qp}$). This has been done by some
authors in the special case where $p=1$ and $q$ arbitrary.

\sb

\centerline {\bf 4. The Algebra $W_{qpQ}$}

We have seen that starting from (1-3), we arrive at
$$
[N,a^+]_1 = a^+ \qquad
[N,a]_1   = -a  \qquad
[a,a^+]_Q = {1 \over {q-p}} [q^N (q-Q) - p^N (p-Q)]
\eqno (35)
$$
(cf.~(10) and (13)).
Reciprocally, we may ask the question~: are there other
hilbertean representations, besides the usual Fock representation
characterized by equations (1-3), of the {\it oscillator algebra}
$W_{qpQ}$ generated by the operators $a$, $a^+ = (a)^{\dagger}$
and $N = (N)^{\dagger}$ satisfying (35) with the assumption that
the spectrum of $N$ is discrete and non-degenerate~? Rideau [47]
has given a positive answer to this question in the special
case where $p=q^{-1}$ and $Q=q$.

We shall not discuss these matters in detail. It is enough to
give the key equation (equation (36) below) that permits to
extend to $W_{qpQ}$ the results valid for $W_q \equiv
W_{qq^{-1}q}$. Indeed, it can be shown that the relations
$$
\eqalign{
a^+ \> |n> \; &= \; \left( Q^{n+1} \lambda_0
+ p^{\nu_0} {{Q^{n+1} - p^{n+1}} \over {q-p}}
+ q^{\nu_0} {{q^{n+1} - Q^{n+1}} \over {q-p}} \right)^{1 \over 2}
\; |n + 1 > \cr
a   \> |n> \; &= \; \left( Q^{n}   \lambda_0
+ p^{\nu_0} {{Q^{n}   - p^{n}}   \over {q-p}}
+ q^{\nu_0} {{q^{n}   - Q^{n}}   \over {q-p}} \right)^{1 \over 2}
\; |n - 1 > \cr
N   \> |n> \; &= \; (\nu_0 + n) \; |n> \cr
}
\eqno (36)
$$
provide us with a {\it formal} representation of $W_{qpQ}$.
Equation (36) leads to three types of representations~:
$T(\nu_0^{\prime}) $ with $n \in \grn$,
$T(\lambda_0,\nu_0)$ with $n \in \grz$ and
$T(\nu_0)          $ with $n \in \grz$. (As a particular case, the Fock
representation (1-3) is obtained from (36) by taking
$\lambda_0 = \nu_0 = 0$ and $n \in \grn$.) These three types of representations
coincide with the ones found by Rideau [47] for $q = p^{-1} = Q$.

\sb

\centerline {\bf 5. From $qp$-Analogues of Angular Momenta to
$u_{qp}(2)$}

We now continue with the Hilbert space
$$
{\cal E} \, = \, \left\{ |jm> \ \; : \; 2j \in \grn, \ m = -j(1)j \right\}
\eqno (37)
$$
spanned by the common eigenvectors of the $z$-component and the
square of a generalized angular momentum. Here, again, we work
with non-deformed state vectors.

Definition 4. We define the
operators $a_+$,    $a_+^+$,
          $a_-$ and $a_-^+$ on the vector space ${\cal E}$ by the relations
$$
\eqalign{
a_+ \; |jm> \; = \; & {\sqrt{[j + m]}}\; |j - {1\over 2}, m - {1\over 2}>\cr
a_+^+ \; |jm> \; = \; & {\sqrt{[j + m+1]}}\; |j + {1\over 2}, m + {1\over
2}>\cr
a_- \; |jm> \; = \; & {\sqrt{[j - m]}}\; |j - {1\over 2}, m + {1\over 2}>\cr
a_-^+ \; |jm> \; = \; & {\sqrt{[j - m + 1]}}\; |j + {1\over 2}, m - {1\over 2}>
\cr
}
\eqno (38)
$$
where the $qp$-numbers of the type $[c]$ are given by (3).

In the limiting case $p = q^{-1} \to 1$, equation (38) gives back
the defining relations used by Schwinger in his (Jordan-Schwinger)
approach to angular momentum (see also Refs.~[21,36]). By introducing
$$
n_1 = j + m \qquad n_2 = j - m \qquad n_1 \in \grn \qquad n_2 \in \grn
\eqno (39)
$$
and
$$
|jm> \, \equiv \, |j + m, j - m> \, = \, |n_1n_2> \; \; \in \;
{\cal F}_1 \otimes {\cal F}_2
\eqno (40)
$$
equation (38) can be rewritten in the form
$$\eqalign{
  a_+   \; |n_1n_2> \; = \; &{\sqrt {[n_1]}}\; |n_1 - 1, n_2>\cr
  a_+^+ \; |n_1n_2> \; = \; &{\sqrt {[n_1 + 1]}}\; |n_1 + 1, n_2>\cr
  a_-   \; |n_1n_2> \; = \; &{\sqrt {[n_2]}}\; |n_1, n_2 - 1>\cr
  a_-^+ \; |n_1n_2> \; = \; &{\sqrt {[n_2 + 1]}}\; |n_1, n_2 + 1>\cr
  }
\eqno (41)
$$
Observe that
$$
\vert jm > = { 1 \over \sqrt{[j+m]! [j-m]!} } \>
(a_+^+)^{j+m} \>
(a_-^+)^{j-m} \> \vert 00 >
\eqno (42)
$$
and, therefore, it is possible to generate any state vector
$\vert jm>$ from the state vector $\vert 00>$
(cf.~equation (8)).

The sets $\left\{ a_+, a_+^+ \right\}$
     and $\left\{ a_-, a_-^+ \right\}$ are two commuting sets of
$qp$-bosons. More precisely, we have
$$
\eqalign{
a_+a^+_+ \; - \; q a^+_+ a_+ \; & = \; p^{N_1} \qquad
a_+a^+_+ \; - \; p a^+_+ a_+ \;   = \; q^{N_1} \cr
a_-a^+_- \; - \; q a^+_- a_- \; & = \; p^{N_2} \qquad
a_-a^+_- \; - \; p a^+_- a_- \;   = \; q^{N_2}
}
\eqno (43)
$$
and
$$
  [a_+, a_-]     \; = \;
  [a^+_+, a^+_-] \; = \;
  [a_+, a^+_-]   \; = \;
  [a^+_+, a_-]   \; = \; 0
\eqno (44)
$$
with
$$N_1 |n_1n_2 > \; = \; n_1 |n_1n_2> \qquad
  N_2 |n_1n_2 > \; = \; n_2 |n_1n_2 >
\eqno (45)
$$
defining the number operators $N_1$ and $N_2$. We also have
$$
a_+a^+_+ = [N_1 + 1] \qquad a^+_+ a_+ = [N_1] \qquad
a_-a^+_- = [N_2 + 1] \qquad a^+_- a_- = [N_2]
\eqno (46)
$$
to be compared with equation (11).

Definition 5. Let us consider the operators
$$
J_- \; = \; a^+_- a_+ \qquad
J_3 \; = \; {1 \over 2} \left( N_1 - N_2 \right) \qquad
J   \; = \; {1 \over 2} \left( N_1 + N_2 \right) \qquad
J_+ \; = \; a^+_+ a_-
\eqno (47)
$$
defined in terms of $qp$-bosons.

Property 5. The action of the linear operators $J_-$, $J_3$,
$J$ and $J_+$ on the space ${\cal E}$ is described by
$$
\eqalign{
  J_- \; |jm > \; = \; &{\sqrt {[j + m] \; [j - m + 1]}} \; |j, m-1 >\cr
  J_3 \; |jm > \; = \; &m \; |jm > \qquad
  J   \; |jm > \; = \;  j \; |jm > \cr
  J_+ \; |jm > \; = \; &{\sqrt {[j - m] \; [j + m + 1]}} \; |j, m+1 >\cr
}
\eqno (48)
$$
a result that follows from (38) and (47).

The operators $J_-$ and $J_+$ are clearly shift operators for
the quantum number $m$. Repeated application of (48) leads to
$$
\eqalign{
\vert jm > & = \left( {[j+m]! \over {[2j]![j-m]!}}
\right)^{1 \over2} \> (J_-)^{j-m} \> \vert j,+j> \cr
\vert jm > & = \left( {[j-m]! \over {[2j]![j+m]!}}
\right)^{1 \over2} \> (J_+)^{j+m} \> \vert j,-j> \cr
}
\eqno (49)
$$
Furthermore, we have the hermitean conjugation
properties~:
$J   = (J  )^{\dagger}$,
$J_3 = (J_3)^{\dagger}$ and
$J_+ = (J_-)^{\dagger}$. Note that $J_-$, $J_3$ and $J_+$
reduce to ordinary spherical angular momentum operators in the
limiting case $p = q^{-1} \to 1$. The latter assertion is evident from
(48) or even directly from (47).

At this stage, the quantum algebra $u_{qp}(2)$ can be introduced,
in a pedestrian way, from equations (47) and (48) as a
deformation of the ordinary Lie algebra of the unitary
group $U(2)$. In this regard, we have the following property.

Property 6. The commutators of the $qp$-deformed
operators $J_-$, $J_3$, $J$ and $J_+$ are
$$
\eqalign{
[J  , J_3] \;&= \; 0        \qquad
[J  , J_+] \; = \; 0        \qquad
[J  , J_-] \; = \; 0        \cr
[J_3, J_-] \;&= \; - \; J_- \qquad
[J_3, J_+] \; = \; + \; J_+ \qquad
[J_+, J_-] \; = \; (qp)^{J - J_3} \> [2J_3]
}
\eqno (50)
$$
which reduce to the familiar expressions, known in the angular
momentum theory, in the limiting case $p = q^{-1} \to 1$.

Note that changing $q$ into $p$ and vice versa does not change
the commutation relations (50). Note also that in the basis
$$
\left  \{ J, \ J_x = (J_+ + J_-)/2,
             \ J_y = (J_+ - J_-)/(2i),
             \ J_z = J_3
\right \}
\eqno (51)
$$
the last three relations in (50) can be rewritten as
$$
[J_x , J_y] = i {1 \over 2} (qp)^{J-J_z} [2J_z] \qquad
[J_y , J_z] = i J_x                             \qquad
[J_z , J_x] = i J_y
\eqno (52)
$$
which exhibit an anisotropy (or symmetry breaking) when
compared to the corresponding relations for the limiting case
$p = q^{-1} \to 1$.

Equation (50) is at the root of the definition of the quantum
algebra $u_{qp}(2)$. Loosely speaking, this algebra is spanned by
any set ${J_-, J_3, J, J_+}$ of four operators satisfying (50) where
we recognize familiar commutators except for the last one.

The case $p = q^{-1}$ deserves a special
attention. In this case, we have
$$
[J_3, J_-] \; = \; - \; J_- \qquad
[J_3, J_+] \; = \; + \; J_+ \qquad
[J_+, J_-] \; = \; [2J_3]
\eqno (53)
$$
where $[2J_3]$ stands here for
$(q^{2J_3} - q^{-2J_3}) / (q - q^{-1})$.
Equations (53) define the quantum algebra $su_q(2)$ addressed by
many authors. It is to be mentioned that there are several
other deformations of $su(2)$ besides the
Kulish-Reshetikhin-Drinfeld-Jimbo deformation [3,6,7] characterized by
equation (53). Let us simply quote, among others, the
(one-parameter) deformation by Woronowicz [8] and the
(one- and two-parameter) deformations by Witten [14] and by Fairlie [16].

In the limiting case $p=q^{-1} \to 1$, the
(infinite dimensional)
quantum algebras $su_q(2)$ and $u_{qp}(2)$ reduce to the
(finite dimensional)
ordinary Lie algebras $su(2)$ and $u(2)$, respectively. We note
that the composition law (i.e., the commutator law) for the
quantum algebras under consideration is anti-symmetrical and
satisfies the Jacobi identity. However, quantities of the type
$[2J_3]$ are not in the vector space generated by $J_-$, $J_3$
and $J_+$. They rather belong to the universal enveloping
algebra of $su(2)$ so that neither $su_{q}(2)$ nor $u_{qp}(2)$
are Lie algebras. (This is at the root of another terminology~:
a {\it quantum algebra} is also referred to as a {\it quantized
universal enveloping algebra}~; in this respect, the
$q$-quantized and $qp$-quantized universal enveloping algebras $su_q(2)$
and $u_{qp}(2)$ may be denoted as $U_q(su(2))$ and
$U_{qp}(u(2))$, respectively.)

Because the algebras $su_q(2)$ and $u_{qp}(2)$ are not Lie algebras,
it is not possible to get Lie groups from them by using the usual
``exponentiation'' procedure. (It is possible, however, to generate
a quantum algebra from what is called a compact matrix pseudo-group by
making use of an unusual ``derivation'' procedure [8].
In the pioneer work by Woronowicz [8], the entries of the matrix
of a pseudo-group are non-commutative objects but usual matrix
multiplication is preserved.) We shall
briefly see in section 6 that the quantum algebras $su_q(2)$
and $u_{qp}(2)$ can be endowed with the structure of a
(quasi-triangular) Hopf algebra.

The notion of invariant operator also exists for
quantum algebras. In this connection, we can verify that the
operator
$$
J^2 \; = \; {1 \over 2} \;
(J_+J_- + J_-J_+) + {{[2]} \over {2}} \; (qp)^{J-J_3} \; [J_3]^2
\eqno (54)
$$
is a Casimir operator in the sense that it commutes with each of
the generators $J_-$, $J_3$, $J$ and $J_+$ of the quantum algebra
$u_{qp}(2)$. Alternatively, we have
$$
J^2 = J_+ J_- + (qp)^{J-J_3+1} [J_3] [J_3 - 1]
    = J_- J_+ + (qp)^{J-J_3}   [J_3] [J_3 + 1]
\eqno (55)
$$
It can be proved that the eigenvalues of the
hermitean operator $J^2$ are
$$
{{q^{2j+1} - q^j p^{j+1} - q^{j+1}p^j + p^{2j+1}} \over {(q - p)^2}}
\equiv [j] \, [j+1]
\eqno (56)
$$
with $2j \in \grn$,
a result compatible with the well-known one corresponding to the
limiting case $p = q^{-1} \to 1$.

Definition 6. We now introduce the operators
$$
K_- \; = \; a_+a_- \qquad
K_3 \; = \; {1 \over 2} \; (N_1 + N_2 + 1)
    \; \equiv \; J + {1 \over 2} \qquad
K_+ \; = \; a^+_+ a^+_-
\eqno (57)
$$
which are indeed $qp$-deformed hyperbolic angular momentum
operators.

Property 7. The action of the operators $K_-$, $K_3$ and $K_+$
on the space ${\cal E}$ is described by
$$
\eqalign{
  K_- \; |jm > \; = \; &{\sqrt {[j-m] \, [j+m]}}     \; |j-1, m >\cr
  K_3 \; |jm > \; = \; &(j+{1 \over 2}) \; |jm >\cr
  K_+ \; |jm > \; = \; &{\sqrt {[j-m+1] \, [j+m+1]}} \; |j+1, m >\cr
}
\eqno (58)
$$
a result to be compared with (48).

The operators $K_-$ and $K_+$ behave like shift operators for
the quantum number $j$. The operators $K_-$,
$K_3 = (K_3)^{\dagger}$ and
$K_+ = (K_-)^{\dagger}$
reduce to ordinary hyperbolic angular momentum operators in the
limiting case $p = q^{-1} \to 1$. From equation (58), we expect that
they generate an algebra that reduces to $su(1,1)$ when
$p = q^{-1} \to 1$.

Property 8. The commutators of the $qp$-deformed
operators $K_-$, $K_3$, $J_3$ and $K_+$ are
$$
\eqalign{
[J_3, K_3] &= 0          \qquad
[J_3, K_+]  = 0          \qquad
[J_3, K_-]  = 0          \cr
[K_3, K_-] &= - K_-      \qquad
[K_3, K_+]  = + K_+      \cr
[K_+, K_-] &= - [2K_3] + (1 -qp)[K_3 + J_3 - {1 \over 2}]
                                [K_3 - J_3 - {1 \over 2}]
}
\eqno (59)
$$
which lead to familiar expressions in the limiting case
$p=q^{-1} \to 1$.

Indeed, equation (59) shows that the operators $K_-$, $K_3$,
$J_3$ and $K_+$ span the ordinary Lie algebra $u(1,1)$ when
$p=q^{-1} \to 1$. When $p=q^{-1}$, with $q$ arbitrary, equation
(59) yields
$$
[K_3, K_-] \; = \; - \; K_- \qquad
[K_3, K_+] \; = \; + \; K_+ \qquad
[K_+, K_-] \; = \; - \; [2K_3]
\eqno (60)
$$
with $[2K_3] = (q^{2K_3} - q^{-2K_3}) / (q - q^{-1})$,
so that $K_-$, $K_3$ and $K_+$ generate the quantum algebra
$su_q(1,1)$ or $U_q(su(1,1))$
worked out by many authors. When $p \ne q^{-1}$,
equation (59) may serve to define the quantum algebra
$u_{qp}(1,1)$ or $U_{qp}(u(1,1))$.

To close this section, let us mention that the $J$'s and the
$K$'s do not close under commutation. However, by introducing
four other bilinear forms in the $a$'s
(viz.,
$k^+_+ = - (a^+_+)^2$,
$k^+_- =   (a^+_-)^2$,
$k^-_+ =   (a_-)  ^2$ and
$k^-_- = - (a_+)  ^2$),
we end up with a
$qp$-deformed algebra that reduces to $so(3,2) \simeq sp(4,\gr)$
in the limiting case $p = q^{-1} \to 1$ (see Refs.~[36,50]).
This fact might be a good starting point for studying
the quantum algebra $so_{qp}(3,2)$ or $U_{qp}(so(3,2))$.

\sb

\centerline {\bf 6. Representation Theory of $u_{qp}(2)$}

We now focus our attention on the algebra $u_{qp}(2)$. We shall
examine in turn the (irreducible) representations of $u_{qp}(2)$
and describe with some details its Hopf algebraic structure.

The matrix elements of the ($qp$-deformed) generators
$J_-$, $J_3$, $J$ and $J_+$ on the subspace
${\cal E} (j) = \left\{ \vert jm > \; \; : \; m = -j(1)j \right\}$
of ${\cal E}$ follow from equation (48). The corresponding matrices
clearly define a representation, noted $(j)$, of the
algebra $u_{qp}(2)$. As a result, this representation is irreducible since
there is no invariant subspace exactly as in the limiting case
where $p = q^{-1} \to 1$. Of course, this result is valid
only when none of the $qp$-deformed numbers occurring in (48) is
vanishing.

Alternatively, the things can be presented as follows. The matrix
elements of the
$qp$-deformed generators $J_-$, $J_3$, $J$ and $J_+$
of $u_{qp}(2)$ can be connected to those of some
$hf$-deformed generators $A_-$, $A_3$, $A$ and $A_+$
acting also on the subspace
${\cal E} (j)$ and spanning the algebra $u_{hf}(2)$.
As a matter of fact, we have
$$
\eqalign{
J_- & = \sigma(J_3,J) \> A_- = A_- \> \sigma(J_3 - 1,J) \cr
J_3 & = A_3 \quad J = A                                 \cr
J_+ & = A_+ \> \sigma(J_3,J) = \sigma(J_3 - 1,J) \> A_+ \cr
}
\eqno (61)
$$
where the scaling factor $\sigma$, that parallels the scaling
factor $s$ for deformed bosons, is given by
$$
\sigma (J_3, J)=\left( [J - J_3]_{qp} [J + J_3 + 1]_{qp} \right)^  {1 \over 2}
             \> \left( [J - J_3]_{hf} [J + J_3 + 1]_{hf} \right)^{-{1 \over 2}}
\eqno (62)
$$
Therefore, the
$qp$-deformed generators of the quantum algebra $u_{qp}(2)$ are simply
related to the ordinary generators of the Lie algebra $u(2)$
which correspond to $f = h^{-1} \to 1$. In this limiting case,
the matrices on the subspace ${\cal E} (j)$ of the
$qp$-deformed generators $J_-$, $J_3$, $J$ and $J_+$
can be deduced from the well-known ones of the
non-deformed generators $A_-$, $A_3$, $A$ and $A_+$. In this
respect, any $qp$-deformation of $u(2)$ is more or less
equivalent to $u(2)$.

As a net result, the representations of $u_{qp}(2)$ may be
constructed in a simple manner~: To each irreducible
representation $(j)$ of $u(2)$ corresponds an irreducible
representation, denoted by $(j)$ too, of $u_{qp}(2)$ (and reciprocally) of the
same dimension and with the same weight spectrum. Note that the
$(0)$ and $({1 \over 2})$ irreducible representations have the
same matrices in $u(2)$ and $u_{qp}(2)$. (For $j = {1 \over 2}$,
the matrices of the $qp$-deformed generators $2J_-$, $2J_3$ and
$2J_+$ coincide with
the Pauli matrices $\sigma_- = \sigma_1 - i \sigma_2$,
                   $\sigma_3$ and
                   $\sigma_+ = \sigma_1 + i \sigma_2$,
respectively.)

The latter result requires that none of the
$qp$-deformed numbers $[n]$, with $n \in \grn$, involved
with (61) and (62), be vanishing.
Otherwise, we easily understand that a given irreducible representation of
$u(2)$ could become reducible when passing from $u(2)$ to $u_{qp}(2)$.
This may happen, for instance, when $q$ is a root of unity in
the case where $p = q^{-1}$. As a trivial example, the reader
may check that for $p^{-1} = q = \exp (i {2 \pi \over 3})$,
the representation (${3 \over 2}$) of the corresponding quantum
algebra $u_{qp}(2)$
coincides with the direct sum
$(0) \oplus ({1 \over 2}) \oplus (0)$
of the irreducible representations
$(0)$, $({1 \over 2})$ and $(0)$ of $su(2)$. (By denoting the
representations by their dimensions,
we have, in a more suggestive way, $4 = 1 \oplus 2 \oplus 1$.)
This example illustrates the fact that we have some
symmetry breaking~: the irreducible representation $({3 \over 2})$
of $u_{qp}(2)$, with generic $p$ and $q$, gives rise to a (completely)
reducible representation in the limiting situation where
$p^{-1} = q = \exp (i {2 \pi \over 3})$.

Since the algebras $su_q(2)$ and $u_{qp}(2)$ are not Lie
algebras, we forsee that the composition (or product) of two
representations of $su_q(2)$ or $u_{qp}(2)$ differs from the
one of two representations of $su(2)$ or $u(2)$, respectively.
As an illustration, it is easy to check that, if the sets
$\left\{ J_{\alpha}(i) \; \; : \; \alpha = {-},3,{+} \right\}$
for $i=1,2$ span $su_q(2)$, then the set
$$
\left\{ \Delta(J_{\alpha}) = J_{\alpha}(1) \otimes I(2) +
I(1) \otimes J_{\alpha}(2) \; \; : \; \alpha = {-},3,{+} \right\}
\eqno (63)
$$
does not span $su_q(2)$ except for $q=1$. (We use $I$ to denote the identity
operator.) In other words, the operation $\Delta$ is not $su_q(2)$
co-variant for $q \ne 1$. In the terminology of angular momentum theory, it
is not possible to construct a resulting $q$-deformed angular momentum
$\bf J$ from the ordinary sum ${\bf J}(1) + {\bf J}(2)$ of two
$q$-deformed angular momenta ${\bf J}(1)$ and ${\bf J}(2)$ when
$q \ne 1$.

A correct way of doing the product of representations (or of
coupling deformed angular momenta) is to introduce the notion
of co-product. In the case of the quantum algebra $u_{qp}(2)$, this
may be achieved as follows.

Definition 7. Let $\Delta_{qp}$ be the algebra
homomorphism defined, for $u_{qp}(2)$, by
$$
\eqalign{
& \Delta_{qp} (J_3)    = J_3 \otimes I + I \otimes J_3 \qquad
\Delta_{qp} (J  )      = J   \otimes I + I \otimes J   \cr
&\Delta_{qp} (J_{\pm}) = J_{\pm} \otimes
  (qp)^{{1 \over 2}J} (qp^{-1})^{+ {1 \over 2}J_3}
+ (qp)^{{1 \over 2}J} (qp^{-1})^{- {1 \over 2}J_3} \otimes J_{\pm} \cr
}
\eqno (64)
$$
where we have abbreviated $X(1) \otimes Y(2)$ by $X \otimes Y$.

Property 9. The $\Delta_{qp}()$'s defined by (64) satisfy
the commutation relations (50) and thus span $u_{qp}(2)$.

The latter property constitutes an important result.
Equation (64) indicates how to couple (or co-multiply) two
representations of $u_{qp}(2)$. For example, in order of
co-multiply two irreducible representations ($j_1$) and
($j_2$) of $u_{qp}(2)$, we have to replace, in the various $X \otimes Y$
occurring in (64), $X$ and $Y$ by their matrices taken the
($j_1$) and ($j_2$) representations, respectively. From a
physical viewpoint, $\Delta_{qp}$ provides us with a
$qp$-analogue for the composition of two $qp$-deformed angular
momenta. In mathematical terms, $\Delta_{qp}$ is called a co-product.
It is co-associative but not co-commutative.

We have the permutational property
$$
\Sigma (\Delta_{qp}) = \Delta_{pq}
\eqno (65)
$$
where $\Sigma$ is the (transposition) operator defined by
$\Sigma (X(1) \otimes Y(2)) = Y(1) \otimes X(2)$. We note in
passing that $\Delta_{qp}$ and its ``double'' $\Delta_{pq}$ are
connected through a matrix $R_{qp}$ in the following way
$$
\Delta_{qp} = R_{qp} \Delta_{pq} (R_{qp})^{-1}
\eqno (66)
$$
In the case $p = q^{-1}$, $R_{qp}$ is known as the universal
$R$-matrix of Drinfeld. (The $R$-matrix leads to solutions of
the quantum Yang-Baxter equation.)

Besides the co-product $\Delta_{qp}$, it is necessary to
introduce a co-unit (an homomorphism) $\varepsilon$ and an
anti-pode (an anti-homomorphism) $S$ in order that the algebras
$su_q(2)$ and $u_{qp}(2)$ become Hopf algebras. We shall not
insist on these two further notions. In the case of $u_{qp}(2)$,
it is sufficient to know that the homomorphism $\varepsilon$ maps the
$J_{\alpha}$'s onto $0$ or $I$ and the anti-homomorphism $S$ is
defined by
$$
S(J_+) = - (qp^{-1})^  {1 \over 2}  \> J_+ \quad
S(J_3) = - J_3                             \quad
S(J)   = + J                               \quad
S(J_-) = - (qp^{-1})^{-{1 \over 2}} \> J_-
\eqno (67)
$$
To be complete, we should examine the
various compatibility relations which
have to be satisfied as a consequence
of the introduction of the co-product,
co-unit and anti-pode (or co-inverse).
We leave this point to the reader.

Let us close with some general remarks about the
Wigner-Racah algebra of $u_{qp}(2)$. In the case where we
couple (via the co-product $\Delta_{qp}$) two irreducible
representations $(j_1)$ and $(j_2)$ of $u_{qp}(2)$, the
resulting representation is in general a (completely) reducible
representation of $u_{qp}(2)$. The reduction is accomplished by
means of $qp$-deformed coupling matrices, the elements of which
are $qp$-deformed Clebsch-Gordan coefficients from which we
can define symmetrized $(3-jm)_{qp}$ symbols. These coupling
coefficients ($qp$-deformed Clebsch-Gordan coefficients or
$(3-jm)_{qp}$ symbols) may then be combined to produce
$qp$-deformed recoupling coefficients (e.g., in the form
of $(6-j)_{qp}$, $(9-j)_{qp}$, $\cdots$ symbols). The algebraic
relations involving $qp$-deformed coupling and recoupling
coefficients of $u_{qp}(2)$
furnish the basis for developing an $u_{qp}(2)$
Wigner-Racah calculus. Indeed, the $qp$-analogue of an
irreducible tensor operator can be defined for $u_{qp}(2)$.
This naturally leads to a $qp$-analogue of the Wigner-Eckart theorem.
Even further, the concept of an $u_{qp}(2)$ unit tensor can be
developed exactly as in the limiting case $p = q^{-1} \to 1$
(see Ref.~[50]).

\sb

\centerline {\bf 7. Towards Applications}

Rather than dealing with a specific application, as done during
the lectures, we shall concentrate in this section on the
general philosophy inherent to {\it some}
applications of quantum algebras to physics. As a preliminary,
a few remarks are in order.

(i) One may first ask the question~: is there anything new with
$q$- or $qp$-deformed objects~? Equations of type (17) and (61)
incline one to think that deformed objects are more or less
equivalent to the corresponding non-deformed objects. On the
other hand, an operator developed in terms of deformed bosons
(e.g., $a a^+ + a^+ a$) generally exhibits a spectrum
that differs from the one corresponding to non-deformed
bosons. The new representations (36) of $W_{qpQ}$ go in the sense
of a positive answer too.

(ii) To every quantum dynamical system, one can associate a
(non-unique) $qp$-deformed partner or $qp$-analogue.
This may be achieved indeed
in several ways. For example, one may $qp$-deform the dynamical
invariance algebra of the system or solve the $qp$-deformed
Schr\"odinger (or Dirac) equation of the system~; there is no
reason to obtain the same result for the energy spectrum in the
two approaches.

(iii) The Lie algebra $su(2)$ enters many fields of theoretical
physics. Therefore, if the quantized universal enveloping
algebra $su_q(2)$ comes to play a role, it is hardly
conceivable to have an {\it universal significance} for the
deformation parameter $q$. In this connection, one may
expect, for example,
that $q$ be a function of the fine structure constant $\alpha$
in quantum electrodynamics and a function of the so-called Weinberg
angle $\theta_W$ in the theory of electroweak interactions.

The reader has understood from (i)-(iii) that a wind of
{\it pessimism} sometimes accompanies the quantum group invasion.
On the other side, there are plenty of avenues of investigation
based on the concept of quantum algebra. In this direction, the
following four series of applications should constitute an
encouragement towards a certain {\it optimism}.

1. A first series of applications merge from equations (2), (32)
and (56). More precisely, in any problem involving ordinary
bosons or ordinary
harmonic oscillators or ordinary angular momenta (any kind of
angular momentum), one may think of replacing them by the
corresponding $qp$-deformed objects. If the limiting case
where $p = q^{-1} \to 1$ describes the problem in a reasonable way,
one may expect that the case where $p$ and $q$ are close
to $1$ can describe some fine structure effects. In this approach,
the (dimensionless) parameters
$p$ and $q$ are two further fitting parameters describing
additional degrees of freedom~; the problem in
this approach is to find a physical interpretation of the
(fine structure or anisotropy or curvature) parameters $p$ and
$q$. Along this first series, we have the following items.

(i)  Application of $q$-deformed
(and, more generally, $qp$-deformed) harmonic oscillators and of $su_{q}(1,1)$
(and, more generally, $u_{qp}(1,1)$) to vibrational spectroscopy
of molecules and nuclei.

(ii) Application of the algebras $su_{q}(2)$ and $u_q(2)$
(and, more generally, $u_{qp}(2)$) to
(vibrational-)rotational spectroscopy of molecules and nuclei.

(iii) Let us also suggest that
$qp$-bosons might be of interest for investigating the
interaction between radiation and matter.

Most of the applications (i) and (ii) have been concerned up to
now with only one parameter (say $q$). The introduction of a
second parameter (say $p$) should permit more flexibility~; this
is especially appealing for rotational spectroscopy of nuclei
that involves two parameters in the VMI (variable moment of
inertia) model.

2. A second series of applications concerns the more general
situation where a physical problem is well described
by a given (simple) Lie algebra $g$. One may then consider to
associate a (one- or) two-parameter
quantized universal enveloping algebra $U_{qp}(g)$ to the Lie
algebra $g$. For
generic $p$ and $q$ (avoiding exotic cases as $p=q^{-1}=$ root
of unity), the representation theory of $U_{qp}(g)$
is connected to the one of $g$ in a trivial manner since we can
associate a representation of $U_{qp}(g)$ to any representation
of $g$ (cf.~$g = u(2)$). Here again, the
case where $p$ and $q$ are close to $1$ may serve to
describe fine structure effects. Symmetries
described by Lie algebras are thus replaced
by quantum algebra symmetries. See for instance the passage
from $g=so(4)$ to $su_q(2) \oplus su_q(2)$ for the hydrogen
atom system [36]. In this example, a $q$-deformation of the
non-relativistic (discrete) energy spectrum leads to results
that parallel those afforded by the Dirac relativistic
equation. (Here, the parameter $q$ can be related to the fine
structure constant $\alpha$.)

3. A third series arises by allowing the deformation parameters
($p$ and $q$) not to be restricted to (real) values close to 1.
Completely unexpected models may result from this approach.
This is the case for instance when $q = p^{-1}$ is a root of
unity for which case the representation theory of a quantum
algebra $U_{qp}(g)$
may be very different from the case of generic $p$ and $q$.
This may be also the case when $p$ and/or $q$ takes (real)
values far from unity.

4. Finally, a fourth series concerns more
fundamental applications (more fundamental in the sense not
being subjected to fitting procedures). We may mention
applications to statistical mechanics, gauge theories,
conformal field theories and so on. Also, quantum groups
(algebras)
might be interesting for a true definition of the quantum
space-time. (See Ref.~[27] for a pertinent discussion.)

\sb

\centerline {\bf 8. Sketch of Bibliography}

It is not the aim of this section to give an exhaustive list of
references. We shall give under the form of a listing some
references of interest for the theory of quantum algebras and
their applications to physics. Not all the applications are
covered by the listed references. The author apologizes for
not quoting other important papers (known or unknown to him)
about the theory of quantum algebras, the chosen applications
and other possible applications. (Any suggestion to
the author will be appreciated.)
As far as the reader is a beginner on quantum algebras, the
references [1-50]
should be a good starting point. The fifty references
(ordered according to the year of publication) may be
roughly classified as follows. (Sorry for the shortcomings
in this tentative classification~!)

Some basic papers on quantum algebras~: [3, 4, 6, 7, 8, 9, 26, 27].
Paradigms of quantum algebras for physicists~:
[16, 19, 25, 27, 41, 42]. One-parameter deformed bosons~: [1,
2, 5] and many other references. Deformed oscillators~: [5, 10,
11, 13, 17, 18, 24]. The $q$-oscillator ($q$-boson) algebra~:
[10, 11, 47, 48]. On $q$-boson realizations of
$su_{q}(n)$ and
$su_{q}(1,1)$~: [10, 11, 12, 17] and many other references.
The quantum algebra $u_q(3)$~: [40]. On $qp$-bosons,
$qp$-oscillators, $su_{qp}(2)$ and $u_{qp}(2)$~: [28, 38, 45,
49, 50]. Wigner-Racah algebra of $su_q(2)$ and $u_{qp}(2)$~:
[11, 20, 21, 39, 49, 50]. Coherent states~: [29, 35]. On the
$q$-deformed Schr\"odinger equation~: [24, 43]. Nuclear
physics~: [14, 22, 33]. Atomic physics~: [36, 43, 46].
Rotational spectroscopy of (diatomic)
molecules [22, 23, 30, 32] and (deformed and super-deformed) nuclei [22, 33].
Vibrational spectroscopy of molecules~: [31, 32, 34, 37].
Spin Heisenberg chain~: [15]. Condensed matter physics~: [44].

\sb

\centerline {\bf Acknowledgments}

The author would like to thank the Organizing Committee and
more especially Prof.~T.~Lulek for the perfect organization
of a so interesting and cross-fertilizing school. He also
acknowledges Prof.~T.~Fack (University Lyon-1) for useful
comments on this paper.

\sb

\centerline {\bf Appendix}

We first begin with some formulas useful for dealing
with $qp$-deformed numbers $[c]_{qp}$. From equation (3), we easily get
$$
[c]_{qp} \; = \; [c]_{pq} \qquad [-c]_{qp} \; = \; - (qp)^{-c} [c]_{qp}
$$
Furthermore, the following relations (with $[ \; ] \equiv [ \; ]_{qp}$)
$$
\eqalign{
       [a + b] & = [a] \, q^b + p^{a} \, [b]               \cr
   [a + b + 1] & = [a + 1] \, [b + 1] - qp \, [a] \, [b]   \cr
  [a] \, [b+c] & = [a+c] \, [b] + (qp)^{b} \, [a-b] \, [c] \cr
[a-b] \, [a+b] & = [a]^2 - (qp)^{a-b} \, [b]^2             \cr
}
$$
hold for arbitrary numbers $a$, $b$ and $c$.

In the case where $n$ is a positive integer, we have
$$
[n] = q^{n-1} + q^{n-2} p + q^{n-3} p^2 + \cdots + q p^{n-2} + p^{n-1}
\qquad n \in \grn - \left \{ 0 \right \}
$$
where the factorial of $[n] \equiv [n]_{qp}$ is defined by (9).
As illustrative examples, we have
$$
[0] = 0                         \qquad
[1] = 1                         \qquad
[2] = q + p                     \qquad
[3] = q^2 + qp + p^2            \qquad
[4] = q^3 + q^2 p + q p^2 + p^3
$$
and
$$
[2] \, [2] = qp     \> [1] + [3]             \qquad
[2] \, [3] = qp     \> [2] + [4]             \qquad
[3] \, [3] = (qp)^2 \> [1] + qp \> [3] + [5]
$$
which is reminiscent of the addition rule for angular momenta.

In the case where $ p = q^{-1} $ with $q$ being a root of unity,
i.e.,
$$
p^{-1} = q = \exp ( {i2\pi {{k_1}\over {k_2}}} ) \qquad
k_1 \in \grn \qquad k_2 \in \grn
$$
we have
$$
[c] = {{\sin(2\pi{{k_1}\over {k_2}}c)}\over
       {\sin(2\pi{{k_1}\over {k_2}})}}
$$
For instance,
$$
k_1 = 1 \quad k_2 = 4 \quad \Rightarrow \quad q = i = {\sqrt {-1}}
\quad \Longrightarrow \quad [0] = [2] = [4] = \cdots = 0
$$
so that $[c] = 0$ can occur for $c \ne 0$.

We might continue and elaborate on $qp$-arithmetics,
$qp$-combinatorics, $qp$-analysis, $qp$-polynomials and so on.\footnote*
{\sevenrm
The $q$-analysis, which corresponds to $p=1$, goes back to the end of the
nineteenth century and was sometimes referred at this time to as
the ``$q$-disease'' (cf.,
another infestation, viz., the ``Gruppenpest'' in the
first quarter of the twentieth century).}
For the purpose of
the present paper, it is sufficient to restrict ourselves to a
few remarks concerning the $qp$-derivative $D_{qp}$.

As an elementary property, we have
$$
D_{qp} \> x^k = [k]_{qp} \> x^{k-1}
$$
Consequently, $\exp_{qp} (x)$ is an
eigenfunction of $D_{qp}$ with the eigenvalue equal to $1$.
More generally, we obtain
$$
D_{qp} \> \exp_{qp} (\lambda x) = \lambda \> \exp_{qp} (\lambda x)
$$
for $\lambda \in \grc$. In the same spirit, observe that
$$
D_{qp} \> f(\lambda x) =
\lambda \> \left( D_{qp} \> f(y) \right)_{y = \lambda x}
$$
Also of interest, we have
$$
D_{qp} \> \left( f(x) g(x) \right) =
      \left( D_{qp} \> f(x) \right) \> g(px) +
f(qx) \left( D_{qp} \> g(x) \right)
$$
Finally, let us mention that for $p=1$ and $q$ arbitrary, the
derivative $D_{qp}$ is sometimes called the Jackson derivative.

\sb

\centerline {\bf References}

\baselineskip = 0.55 true cm

\noindent
\item{[1]} Arik, M. and Coon, D.D., J.~Math.~Phys.~{\bf 17} (1976) 524.

\noindent
\item{[2]} Kuryshkin, V., Ann.~Fond.~Louis de Broglie {\bf 5} (1980) 111.

\noindent
\item{[3]} Kulish, P.P. and Reshetikhin, N.Yu., Zap.~Sem.~LOMI
{\bf 101} (1981) 101 [J.~Soviet.~Math.~{\bf 23} (1983) 2435].

\noindent
\item{[4]} Sklyanin, E.K., Funkt. Anal. Pril. {\bf 16} (1982) 27
[Funct. Anal. Appl. {\bf 16} (1982) 262].

\noindent
\item{[5]} Jannussis, A., Brodimas, G., Sourlas, D., Vlachos,
K., Siafarikas, P. and Papaloucas, L., Hadronic Journal {\bf 6} (1983) 1653.

\noindent
\item{[6]} Drinfel'd, V.G., DAN SSSR {\bf 283} (1985) 1060
[Soviet Math. Dokl. {\bf 32} (1985) 254]~;
in {\it Proc. Int. Congr. of Mathematics}, ed. A.M.
Gleason, AMS~: Providence (1986).


\noindent
\item{[7]} Jimbo, M.,  Lett. Math. Phys. {\bf 10} (1985) 63~;
                                         {\bf 11} (1986) 247~;
Commun. Math. Phys. {\bf 102} (1986) 537.


\noindent
\item{[8]} Woronowicz, S.L., Publ. RIMS-Kyoto {\bf 23} (1987) 117~;
Commun. Math. Phys. {\bf 111} (1987) 613.

\noindent
\item{[9]} Rosso, M., C.~R.~Acad.~Sc.~Paris {\bf 304} (1987) 323.



\noindent
\item{[10]} Macfarlane, A.J., J.~Phys.~A {\bf 22} (1989) 4581.

\noindent
\item{[11]} Biedenharn, L.C., J.~Phys.~A {\bf 22} (1989) L873.

\noindent
\item{[12]} Sun, C.-P. and Fu, H.-C., J.~Phys.~A {\bf 22} (1989) L983.

\noindent
\item{[13]} Kulish, P.P. and Reshetikhin, N.Yu.,
Lett.~Math.~Phys.~{\bf 18} (1989) 143.


\noindent
\item{[14]} Witten, E., Nucl.~Phys.~B {\bf 330} (1990) 285.

\noindent
\item{[15]} Batchelor, M.T., Mezincescu, L.,
Nepomechie, R.I. and Rittenberg, V.,
J. Phys. A {\bf 23} (1990) L141.

\noindent
\item{[16]} Fairlie, D.B., J.~Phys.~A {\bf 23} (1990) L183.

\noindent
\item{[17]} Kulish, P.P. and Damaskinsky, E.V.,
J.~Phys.~A {\bf 23} (1990) L415.

\noindent
\item{[18]} Chaichian, M. and Kulish, P., Phys. Lett. {\bf 234B} (1990) 72.

\noindent
\item{[19]} Curtright, T.L. and Zachos, C.K.,
Phys.~Lett.~{\bf 243B} (1990) 237.

\noindent
\item{[20]} Biedenharn, L.C. and Tarlini, M.,
Lett.~Math.~Phys.~{\bf 20} (1990) 271.

\noindent
\item{[21]} Nomura, M., J.~Phys.~Soc.~Jpn.~{\bf 59} (1990) 1954 and 2345.

\noindent
\item{[22]} Iwao, S., Prog.~Theor.~Phys.~{\bf 83} (1990) 363.

\noindent
\item{[23]} Bonatsos, D., Raychev, P.P., Roussev, R.P. and
Smirnov, Yu.F., Chem. Phys. Lett. {\bf 175} (1990) 300.

\noindent
\item{[24]} Minahan, J.A., Mod.~Phys.~Lett.~A {\bf 5} (1990) 2625.

\noindent
\item{[25]} Zachos, C.K., in {\it Proc.~of the Spring Workshop
on Quantum Groups}, eds. T.L. Curtright, D.B. Fairlie and C.K.
Zachos, World Scientific~: Singapore (1990).

\noindent
\item{[26]} Woronowicz, S.L., Lett.~Math.~Phys.~{\bf 21} (1991) 35.

\noindent
\item{[27]} Flato, M. and Lu, Z.C., Lett.~Math.~Phys.~{\bf 21}
(1991) 85. See also~: Flato, M. and Sternheimer, D.,
Lett.~Math.~Phys.~{\bf 22} (1991) 155.

\noindent
\item{[28]} Schirrmacher, A., Wess, J. and Zumino, B.,
Z.~Phys.~C {\bf 49} (1991) 317.

\noindent
\item{[29]} Quesne, C., Phys.~Lett.~{\bf 153A} (1991) 303.

\noindent
\item{[30]} Chang, Z. and Yan, H., Phys.~Lett.~{\bf 154A} (1991) 254.

\noindent
\item{[31]} Chang, Z., Guo, H.-Y. and Yan, H.,
Phys.~Lett.~{\bf 156A} (1991) 192.

\noindent
\item{[32]} Chang, Z. and Yan, H., Phys.~Lett.~{\bf 158A} (1991) 242.

\noindent
\item{[33]} Bonatsos, D., Drenska, S.B., Raychev, P.P.,
Roussev, R.P. and Smirnov, Yu.F., J.~Phys.~G {\bf 17} (1991) L67.

\noindent
\item{[34]} Bonatsos, D., Raychev, P.P. and Faessler, A.,
Chem.~Phys.~Lett.~{\bf 178} (1991) 221.

\noindent
\item{[35]} Katriel, J. and Solomon, A.I., J.~Phys.~A {\bf 24} (1991) 2093.

\noindent
\item{[36]} Kibler, M. and N\'egadi, T., J.~Phys.~A {\bf 24} (1991) 5283.

\noindent
\item{[37]} Bonatsos, D., Argyres, E.N. and Raychev, P.P.,
J.~Phys.~A {\bf 24} (1991) L403.

\noindent
\item{[38]} Chakrabarti, R. and Jagannathan, R.,
J.~Phys.~A {\bf 24} (1991) L711.

\noindent
\item{[39]} Smirnov, Yu.F., Tolstoy, V.N. and Kharitonov, Yu.I.,
Sov.~J.~Nucl.~Phys.~{\bf 53} (1991) 959.

\noindent
\item{[40]} Smirnov, Yu.F., Tolstoy, V.N. and Kharitonov, Yu.I.,
Sov.~J.~Nucl.~Phys.~{\bf 54} (1991) 721.

\noindent
\item{[41]} Curtright, T.L., Ghandour, G.I. and Zachos, C.K.,
J.~Math.~Phys.~{\bf 32} (1991) 676.

\noindent
\item{[42]} Cosmas, Z., in {\it Symmetries in Sciences V},
ed.~B.~Gruber, Plenum~: New York (1991).

\noindent
\item{[43]} Xing-Chang Song and Li Liao, J.~Phys.~A {\bf 25} (1992) 623.

\noindent
\item{[44]} Tuszy\'nski, J.A. and Kibler, M., J.~Phys.~A {\bf 25} (1992) 2425.

\noindent
\item{[45]} Katriel, J. and Kibler, M., J.~Phys.~A {\bf 25} (1992) 2683.

\noindent
\item{[46]} N\'egadi, T. and Kibler, M., J.~Phys.~A {\bf 25} (1992) L157.

\noindent
\item{[47]} Rideau, G., Lett.~Math.~Phys.~{\bf 24} (1992) 147~;
Kibler, M. and Rideau, G., unpublished results (1992).

\noindent
\item{[48]} Van der Jeugt, J., Lett.~Math.~Phys.~{\bf 24} (1992) 267.

\noindent
\item{[49]} Wehrhahn, R.F. and Smirnov, Yu.F., Preprint DESY 92-024 (1992).

\noindent
\item{[50]} Smirnov, Yu.F. and Kibler, M., to be published.

\bye